\documentclass[11pt]{article}  
\usepackage[margin=1.2in]{geometry}
\usepackage{graphicx}
\usepackage{float, bm}
\usepackage{amsmath,amssymb,calc,color}
\usepackage{graphicx}
\usepackage{natbib}
\usepackage{bm}
\usepackage{lmodern}

\usepackage{xskak,chessboard}

\newcommand{\defeq}{:=}
\newcommand{\bx}{x}

\newcommand{\mb}[1]{{\bf{#1}}}
\newcommand{\E}{\mathbb{E}}
\newcommand{\I}{\mathbb{I}}

\title{Counting $N$ Queens}
\author{Nick Polson\footnote{ngp@chicagobooth.edu}\\
\textit{Booth School of Business}\\
\textit{University of Chicago}
\and
Vadim Sokolov\footnote{vsokolov@gmu.edu}\\
\textit{Operations Research}\\
\textit{George Mason University}
}

\graphicspath{{fig/}}

\begin{document}
\maketitle
\begin{abstract}
\noindent 
Gauss proposed the problem of how to enumerate the number of solutions for placing $N$ queens on an $N\times N$ chess board, so no two queens attack each other.
The N-queen problem is a classic problem in combinatorics. We describe a variety of  Monte Carlo (MC) methods for counting the number of solutions. In particular, we propose a quantile re-ordering based on the Lorenz curve of a sum that is  related to counting the number of solutions. We show his  approach leads to an efficient polynomial-time solution. Other MC methods include vertical likelihood Monte Carlo, importance sampling, slice sampling, simulated annealing, energy-level sampling, and  nested-sampling. Sampling binary matrices that identify the locations of the queens on the board can be done with a Swendsen-Wang style algorithm.  Our Monte Carlo approach  counts the number of solutions in polynomial time. 
\end{abstract}


\newpage

\section{Introduction}
Gauss introduced the N-queens problem in 1850. The objective is to position $n$ queens on an $n \times n$ chess board such that no one queen can be taken by any other. 
The N-queen problem is a classic problem in combinatorics. It is a generalization of the 8-queen problem, which asks for the number of ways to place 8 queens on a $8 \times 8$ chess board so that no two queens attack each other. The 8-queen problem has 92 solutions. 
This follows from the fact that a traditional chess board with $n=8 $ we have $=12$ configurations. If we consider the rotations, then the number of configurations become $12\times 8 - 1\times 4 = 92$ as 4 of the configurations are identical.  Thus, we have $8$ possible rotations for all the 12 solutions.

\begin{figure}[H]
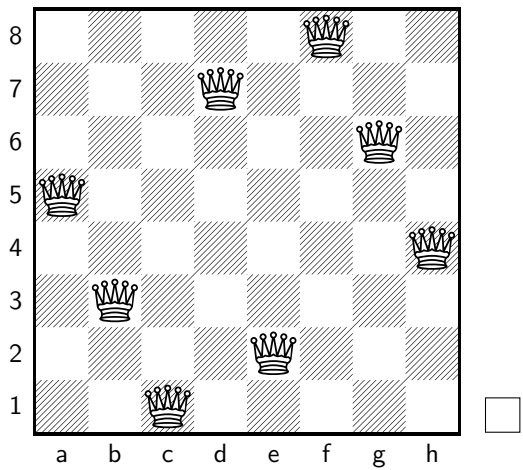
\label{fig:8queens}
  \centering
  \newgame
  \def\myfen{5Q62/3Q4/6Q1/Q7/7Q/1Q6/4Q3/2Q5}
  \newchessgame[setfen=\myfen,moveid=16w]
  \chessboard    
  \caption{A solution to the $8$-Queens problem}
\end{figure}

The position of the Queens can be represented as a binary matrix. Thus, we can view the problem of counting the number of solutions as a problem of sampling from binary Markov random fields.  Specifically,  the position of the queens can be represented as $x\in \{1,\ldots,n\}^n$, where $x_i$ is the column of the queen on the $i$th row. Figure \ref{fig:8queens} shows the $\{6,4,7,1,8,2,5,3\}$ configuration. Naive Monte Carlo samples $x_i$ independently of $\{1,\ldots,n\}$ for $i=1,\ldots,n$. Then the objective function $S$ could be chosen as the number of pairs of queens that can attack each other. The problem with this approach is that the probability of a correct configuration is very small. 

The problem with naive Monte Carlo (MC) is that the configurations represent rare events and thus, are hard to sample using naive approaches. For example, when $n$=9, there 
are $92$ correct configurations out of $4,426,165,368$ possible configurations, making the ratio to be $2.0785\times 10^. {-8}$. This ratio decreases as $n$ goes up. 
The N-queen problem has been solved exactly for $n$ up to 15 and estimated by Monte Carlo for $n$ between 16 and 25. \cite{simkin2022} showed that there exists a constant $\alpha=1.942\pm3\times10^{-3}$ such that the number of configuration sis $((1\pm o(1))ne^{-\alpha})n$. Thus, for $n=25$ there are approximately $10^{15}$ configurations and $10^{31560}$ for $n$ = 10000. Counting \# $N$-Queens can be recast as a rare event expectation calculation. 

 $N$-Queens was  previously analyzed by \cite{higdon1998}, \cite{swendsen1987,polson1996,rubinstein2016}. Sampling from binary random Markov field can be done in polynomial time using the algorithm that was first proposed by \cite{swendsen1987}. The first key insight is to replace the counting problem by a problem of calculating the probability of a rare event \cite{diaconis1995}. Second, we need to find a fast algorithm to sample from a distribution, which can be done in polynomial time due to \cite{swendsen1987}. Third, to reorder the sum as the sum involving the quantile importance sampling. It is well known, that naive MCMC or the product estimator is not polynomial. Thus, we must do the quantile reordering to achieve the polynomial time.

A related important problem is the $N$-Queens completion problem and other approach can be extended to this class of problems. Given an allocation of queens can it be extended to a solution. This is related to the classic $P=NP$ problem as it is known to the \#NP-hard  If an efficient solution exists for one \#NP-hard problem, it can be used for all! Our class of algorithms provides an avenue to finding RP-time algorithms.

The rest of the paper is outlined as follows. Section \ref{sec:MC} describes Monte Carlo counting methods. Including naive MC and sequential MC methods. By constructing an equivalent rare event expectation problem. Section \ref{sec:energy} provides energy domain methods which increase efficiency. Finally, Section \ref{sec:discussion} concludes with comments on the completion problem and the relationship to the $P=NP$ problem.

\section{Monte Carlo MC Counting}\label{sec:MC}
Let $ x \in \mathcal{X} $ be an element in the set of binary matrices. A matrix describes a queen placement. 
The number of solutions will be counted as a rare event probability problem due to \cite{rubinstein2013fast}. Let
$$
S(x) = 1 \; \; {\rm if} \; x \; {\rm solution}  \; \; {\rm and} \; \; S(x) = 0 \; \; {\rm if} \; x \; {\rm not \; a \; solution}
$$
The key insight is the fact that the counting problem can be solved via estimating the rare event probability
\[
\mathbb{P} \left ( \# solns \right ) = E_p \left[ \mathbb{I}_{ S(x) > 0 } \right]=\frac{  \# solns  }{  \# matrices },
\]
where $ p(x ) $ denotes the uniform distribution over the set of binary placement matrices.
This allows as to count the number of possible solutions as we can count the number of possible matrices analytically.

The key insight is that counting is computationally equivalent to a rare event probability problem 
For example, in the $8$-queens problem rather than counting we can estimate using naive MC, the probability 
$$ 
\rho = P( S( x ) = 0 )   = 92/ 4,426,165,368 = 2.0785 \times 10^{-8} 
$$
where here $ S(x) $ can be  defined as the number of attacking positions of the queens. 

When $N=8$ there are $ 4,426,165,368 $ possible binary matrices and this provides the denominator.
The problem now is how to calculate $ \hat{\rho} $ in  random polynomial RP-time! Naive MC will not provide such a method. 

When the counting problem can be embedded into a sub-problem we can save on the efficient computation. For example the $N$-queen problem is a sun-problem of
the $N$-rook problem and there are far fewer $ 8^8 = 16,777,216 $ possible allocations for the denominator. 
One then samples a permutation of the Queens
with $ 8! = 40, 320 $ possible permutations and simply counts the number of solutions, This brute force approach will not provide a polynomial time algorithm for general $N$.

\subsection{Product Estimator}
The issue then becomes how to calculate the rare event probability $ E_p \left[ \mathbb{I}_{ S(x) > 0 } \right] $ \citep{diaconis1995}?  We will do this by a carefully chosen sequential Monte Carlo algorithm. The classic Monte Carlo scheme uses random samples $ x^{(i)} , 1 \leq i \leq N $  
$$
\hat{Z} = \frac{1}{N} \sum_{i=1}^N S( x^{(i)} ) 
$$
The convergence of the variance of $ \hat{Z} $ is $ O( N^{-1} ) $. 

It is more efficient to use a sequential procedure. Define 
$$
Z_\gamma = E_p \left[ \mathbb{I}_{ S(x) >  \gamma } \right]  \; \; {\rm where} \; \; 0 \leq \gamma \leq 1 .
$$
Then, imagine an annealing schedule $ 0 \defeq \gamma_0 < \gamma_1 < \ldots < \gamma_M < 1 $.
This allows us to write the quantity of interest, with  $ Z_1 \defeq 1 $, and 
$$
Z \defeq  Z_0 = E_p \left[ \mathbb{I}_{ S(x) > 0 } \right]  = \prod_\gamma \frac{ Z (\gamma_i) }{ Z ( \gamma_{i-1}) } Z_1  .
$$
The key is to (sequentially) find an optimal annealing schedule so that each of the ratios $ Z (\gamma_i ) / Z ( \gamma_{i-1} )  $ can be calculated in polynomial time.

Now one sequentially samples from  $ \pi ( x |  S(x) > m_{t-1} ) $ and counts how many samples satisfy $ S(x) >m_t$. 
$$
\rho = \mathbb{P} \left ( S(x) > m_t |  S(x) > m_{t-1} \right ) = \mathbb{P} \left ( S(x) > m_t \right ) \mathbb{P} \left (   S(x) > m_{t-1} \right ) = Z_{m_t} / Z_{m_{t-1}} 
$$
Independent samples gives relative mean squared error 
$$
\hat{Z} = \prod_{t=1}^T \widehat{ \frac{Z_{m_t}}{Z_{m_t-1}} } \; \; {\rm and} \; \;
\frac{Var \left ( \hat{Z} \right )}{\mathbb{E} ( \hat{Z} )^2} =  \prod_{t=1}^T
 \left ( \frac{ \hat{\sigma}_{m_t}^2 }{ \hat{\mu}_{m_t}^2 } + 1 \right ) -1
$$
with mean $  \mathbb{E} ( \hat{Z}_{m_t} / \hat{Z}_{m_{t-1}} ) = \hat{\mu}_{m_t} $ and variance $ \hat{\sigma}_{m_t}^2  $.

The key is to pick i $ m_t $ so coefficients of variation are all equal  reduces variance.
A  ``well-balanced'' cooling schedule, with 
Chebyshev-Hoeffding bounds, leads to  a running time of $O \left ( (\ln (1/Z) )^2 \right ) $ steps. However, this is not RP.
See also the cross-entropy method of Rubenstein. 

\paragraph{Cross-Entropy} CE \citep{rubinstein2016} methods solve the problem of estimating
\[
Z = P_h(S(X)\ge \gamma) = E{I(S(X)\ge \gamma)}
\]
Here $X \sim h_{\theta}(X)$ and the parameters $\theta$ of the sampling distribution $h$ and the temperature parameter $\gamma$ are estimated sequential and jointly. The CE method estimates optimal $\theta^*$. This is achieved by  finding an importance sampling density $g$ that is close to the optimal one
\[
g^*(x) = \dfrac{f(x;u)I(S(X)\ge \gamma)}{Z}.
\]
The closeness in measured uins KL divergence
\[
D(g^*,g) = \E{\log \dfrac{g^*(X)}{g(X)}} = \int g^*(x)\log \dfrac{g^*(X)}{g(X)} dx.
\]
\cite{rubinstein2016} (p.211) use a different embedding and use a combinatorial optimisation method to find solutions using a Gibbs sampler. This requires exchanging queens on the chess board. They define  $S(x)$ to represent the amount of ``threat" of the queens. They simply add the number of queens minus one for each row and column. An optimal solution corresponds to the minimum of $S$ namely zero.
Our approach directly counts the number of solutions and provides theoretical methods to address the random polynomial time nature of our algorithm.

If the function $S(x)$ counts the number of solutions to the N-Queens problem, we can use the Boltzman distribution
\[
\pi_{\gamma}(x) = \dfrac{1}{Z(\gamma)}\exp(-S(x)/\gamma).
\]
As $\gamma \rightarrow \infty$, $\pi_{\infty}$ collapses to uniform distribution on the set of solutions $x_{\infty} = \{x \in X ~:~ \arg\min_x S(x)\}$. Thus, the size of the set $x_{\infty}$ counts the number of solutions. 

\subsection{Swendsen-Wang sampling}

Another problem is to count the number of the correct solutions. This problem we can solve by sampling configurations $x\in \{1,\ldots,n\}^n$ where 
$p_i$ is uniform on the column $x_i$ from 
\[
p(x) = \prod_{i=1}^{n}p_i(x_i),
\]
\cite{swendsen1987}  use a latent variable and introduce slice variables $u_i$ and joint distribution 
\[
p(x,u) = \prod_{i=1}^{n}I(0 < u_i \le p_i(x_i)),
\]
The Swendsen-Wang sampling is a variation of the slice sampling algorithm \cite{higdon1998}. We will take advantage of the ``Fortuin-Kasteleyn-Swendsen-Wang'' (FKSW) joint distribution identified explicitly in  \cite{edwards1988} over color variables $s$ and a bond variable for each edge $E$ in the graph.  

The joint distribution is
\[
  P(s,d) = \dfrac{1}{Z}\prod_{(i,j)\in E} \left((1-p)\delta_{d_{ij,0}}+p\delta_{d_{ij,1}}\delta_{d_{s_i,s_j}}\right).
\]
The marginal distribution over $s$ in the FKSW model is the Potts distribution. 
\[
  p(s) \propto e^H, ~ H = -J_p \sum_{(i,j)}\delta(s_i,s_j).
\]
The marginal distribution over the bonds is the random cluster model of \cite{fortuin1972}:
\[
  P(d) = \dfrac{1}{Z}p^D(1-p)^{|E|-D}q^{C(d)} = \dfrac{1}{Z}\exp{\left(D\log\left(d^J-1\right)\right)}e^{-J|E|}q^{C(d)},
\]
where $C(d)$ is the number of connected components in a graph with edges wherever $d_{ij} =1$, and $D= \sum_{(ij)\in E}d_{ij}$.

The algorithm of \cite{swendsen1987} performs block Gibbs sampling on the joint model by alternately sampling from $P(d_{ij}\mid s)$ and $P(s\mid d_{ij})$. This can convert a sample from any of the three distributions into a sample from one of the others.

\paragraph{Clustering Algorithm}
Pick $i,j$, choose symbols $a, b$.
Then scan row $i$ and choose column $j$ where symbol $s_{ij}$ is $a$.
Search other rows for the symbol $b$ at the same column
For each row $i$ use bond variable $u$. The joint distribution is 
\[
  \pi(x) \propto \pi_0(x) \prod_k e^{- \frac{1}{\gamma} S_k(x)}
\]
marginal from joint
\[
  \pi(x,u) \propto \pi_0(x) \prod_k I\left(0<u_k<e^{- \frac{1}{\gamma} S_k(x)}\right)
\]
See also plit sampling \cite{birge2013} and vertical likelihood Monte Carlo \cite{polson2015} for optimal importance sampling choice. 

\cite{rubinstein2016} and \cite{skilling2006} nested sampling uses cross-entropy to sample (not optimal) - and empirical quantiles of  the Boltzmann distribution. This is a family of distributions defined as
\[
 \pi_\gamma ( x ) =   \frac{ e^{- \frac{1}{\gamma} S(x)} } { \sum_{ x \in \mathcal{X}  }  e^{- \frac{1}{\gamma} S(x)}  } 
\]
Picking  $\gamma$ can be done by using a discrete grid or by using an adaptive schedule. 
\cite{masry1990} provide a trapezoidal rule.

\section{Energy-domain counting}\label{sec:energy}

\subsection{Vertical Likelihood Sampling} 

In the vertical likelihood approach \cite{polson2015}, we express $Z$ in a continuous fashion and use importance sampling
\begin{align*}
Z &=  E_p \left[ \mathbb{I}_{ S(x) > 0 } \right]   = \int_0^1 Z( \gamma ) d \gamma  = \int_0^1 Z( \gamma ) \frac{1}{\Omega(\gamma)} \Omega (\gamma ) d \gamma
\end{align*}
The key is that $\Omega (\gamma )$ can be chosen in an optimal fashion.

This can be performed the simulation in a sequential fashion with an increasing schedule as $ \gamma $ goes from $0$ to $1$ as the simulation proceeds.
The key is to do this in polynomial time.

Some methods use fixed schemes for $ \Omega (\gamma) $. For example, nested sampling uses $\gamma_i = e^{- i / N } $ but there is  no guarantee that this is  polynomial time. 

Polson and Scott show that many of the commonly used SMC methods simply correspond to difference choices of $ \Omega ( \gamma ) $.
We can think of $ \Omega (\gamma ) $  as providing the optimal annealing schedule.  How quickly can you increase $ \gamma$. Phase change when you increase $ \gamma $
and the number of solutions explodes exponentially. Hence, you have to be able to control this. 

Examples of energy-domain features include the micro-canonical distribution over the equi-energy surface $\{x: h(x) = s\}$, as well as the microcanonical average of a state function,
$$
\nu_f(s) = \E \{S(x) \mid h(x) = s\} \, ,
$$
which is independent of temperature.  Another energy-domain feature is the density of states.  In the discrete case, this is the number of states with a given energy level: $N(s) = \#\{x: h(x) = s \}$.  In the case where $\mathcal{X}$ is a continuous state space, $N(s)$ is the function such that the volume of the set $\{x: h(x) \in (s, s + ds) \}$ is approximately $N(s) ds$.

The statistical-mechanics community has developed a wide class of Monte Carlo methods to approximate energy-domain features.  We refer to these collectively as energy-level samplers.  These methods all share the goal of biasing the draws toward higher-energy states by means of an iterative re-balancing scheme.

To motivate these methods, let $X$ be a draw from the Boltzmann distribution, assuming $T=1$ without loss of generality.   Let $\eta = h(X)$ be the corresponding random energy level, with distribution
\begin{equation}
\label{eqn:energyleveldensity1}
P(\eta = s) = \sum_{x: h(x) = s } e^{-s} = e^{-s} \ N(s) \, .
\end{equation}
The multi-canonical sampler of \citet{berg1991} attempts to re-balance the sampler so that the implied energy distribution becomes flat: $P(\eta = s) \propto \mbox{constant}$.  As (\ref{eqn:energyleveldensity1}) suggests, this is accomplished by sampling states $x$ with weight inversely proportional to the density of states $N(s)$.

If the density of states is unknown, the Wang--Landau algorithm \citep{wang2001,wang2001a} provides a suitable variation.  It involves estimating $N(s)$ via an iterative re-balancing approach, and has been generalized to a wider class of statistical problems \citep{bornn2013}.  For a short overview of the adaptive Wang--Landau algorithm, see the Appendix.

An even more extreme re-balancing is the $1/k$-ensemble sampler \citep{hesselbo1995}.  Let
\begin{equation}
\label{eqn:cumulativedensitystates}
Z(s) = \#\{x: h(x) \leq s\} = \sum_{t \leq s} N(t)
\end{equation}
define the cumulative number of states with energy as least as small as $s$.  In the $1/k$-ensemble sampler, states are sampled with weight proportional to $Z(s)$, rather than $N(s)$ as in the multi-canonical sampler.  This makes it even easier for the sampler to traverse high-energy (low-probability) regions of the state space.

\paragraph{The connection with latent-variable methods.}  Here we note a connection between auxiliary-variable methods and energy-level samplers that underlies our recommended approach for choosing an importance function.  Motivated by the latent-variable scheme at the heart of slice sampling \citep{damlen1999a}, consider the joint distribution
\begin{equation}
\label{eqn:jointenergy}
p(x,u) \propto w(u) \ \I\{ u \geq h(x) \} \, .
\end{equation}
Let $(X,U)$ be a random draw from this joint distribution, and just as above, consider the implied distribution over the random energy level $h(X)$:
$$
p(h(X) = s) \propto \sum_{x: h(x) = s} w(s) = w(s) N(s) \, .
$$
If $w(s) = e^{-s}$, we recover the canonical ensemble: that is, the distribution over the energy level implied by the original Boltzmann distribution (\ref{eqn:energyleveldensity1}).  On the other hand, if we set $w(s) = 1/N(s)$, we see that $P(h(X) = s)$ is now constant in $s$, as in the multicanonical sampler and Wang--Landau algorithm.  Finally, if we set $w(s) = 1/Z(s)$ as in (\ref{eqn:cumulativedensitystates}), then we obtain
$$
P(h(X) = s) \propto N(s)/Z(s) \, ,
$$
as in the $1/k$-ensemble sampler of \citet{hesselbo1995}.

To summarize: many different sampling schemes historically used for energy-domain features can be interpreted as different choices for the weight function in a joint distribution defined via an auxiliary variable (\ref{eqn:jointenergy}).  The particular form of this joint distribution suggests an interesting connection between slice sampling and energy-level sampling that can be usefully exploited.

\subsection{Quantile Re-ordering }

We can compute $ Z = \mathbb{E}(X) =.  \int_{\mathcal{X}} L(x) p(d x) $ In many cases, we wish to calculate expectations: 
$ Z=E( X) = \sum_{ x \in \mathcal{X} }  L(x ) p( x ) $ where  $L(x)$ is typically in conflict with  $p(x)$.
It is useful to re-order this sum in terms of quantiles before trying to enumerate with Monte Carlo.

Using the mean identity for a positive random variable and its CDF or equivalently, via the Lorenz curve, we obtain the key identities  
\begin{align*}
 E(X) &= \int_{\mathcal{X}} L(x) p(d x) = \int_0^\infty (1 - F_X(x) )d x = \int_0^\infty Z(u)du  \\
E(X) &= \int_0^1 F_X^{-1}(s) ds =  \int_0^1 \Lambda (1- s ) ds = \int_0^1 \Lambda(s) ds  
\end{align*}
We do not have to assume that $ F^{-1}(s)$, or equivalently $ \Lambda (s) $, is available
in closed form, rather we can find an unbiased estimate of this by simulating the Lorenz curve.

Lorenz curve of the likelihood ordinate $ Y \equiv L(X) $ where 
$ X \sim p(x) $. The Lorenz curve, $ \mathcal{L} $ of $X$ is defined in terms
of its CDF, $F_X(x)$, as 
\begin{align*}
  \mathcal{L}(u) & = \frac{1}{Z} \int_0^u F_X^{-1} ( s ) d s \; \; \text{ where } \; u \in [0,1] \\
  Z & = \mathbb{E} ( X) = \int_{ \mathcal{X} } L(x) p(dx ) \; .
\end{align*}
One feature of a Lorenz curve is that is provides a way to evaluate 
$$
 \mathbb{E}(X) = \int_{\mathcal{X}} L(x)p(dx) = \int_0^1 F_X^{-1}(s) d s 
 $$
Note that  under the transformation $ u=F_X(x) $, we have 
$$ E(X) = \int_0^1 F_X^{-1} ( u ) d u  = \int_0^\infty x f_X(x) d x 
$$
Here we use Lebesgue rather than Riemann integration.   \cite{yakowitz1978weighted} shows that the MC standard errors are lower for the quantile approach.

The pseudo-inverse of $Z(y)$, denoted $\Lambda(s)$ and defined as
\begin{equation}
\label{eqn:LambdaZ}
\Lambda(s) = \sup\{y: Z(y) > s\} \, ,
\end{equation}
which, like $Z(y)$, is non-increasing.

Intuitively, $\Lambda(s)$ gives the value $y$ such that $s$ is the fraction of prior draws with likelihood values larger than $y$.
\begin{equation}
Z = \int_{\mathcal{X}} L(x) \ d P(x) =  \int_{\mathcal{X}} \int_0^{\infty}  I\{y < L(x)\}  \ dy \ d P(x) = \int_0^{\infty} Z(y) \ dy  \label{eqn:Zin1D} \, 
\end{equation}
Now, exploiting the fact that $s < Z(y)$ if and only if $y < \Lambda(s)$:
\begin{equation}
Z =  \int_0^{\infty} \int_0^1 \mathbb{I}\{ s < Z(y) \} \ ds \ dy 
=  \int_0^{\infty} \int_0^1 \mathbb{I}\{ y < \Lambda(s) \} \ ds \ dy =   \int_0^1 \Lambda(s) \ ds \label{eqn:Zin1D2} \, .
\end{equation}

We start with the Lorenz identity, stated as a lemma, that is at the heart of nested sampling and the quantile importance sampling.

Since $X \sim  P $, and $ U \sim  U (0, 1)$, it follows that  $\Lambda (U ) \stackrel{D}{=}  L(X)$, that is the likelihood ordinates are distributionally the  same as the $\Lambda(u)$ values at uniform grid points. It also follows from the definition of $ Z$, that $Z ( L(X)  )  \stackrel{D}{=}   U(0, 1)$.
Let us consider the slightly modified alternative presentation of nested sampling by Polson and Scott (2014). 

The evidence 
$ Z  = \int_0^1 \Lambda (s) ds$  is approximated at grid points
$(s_1 > s_2 >  \ldots > s_n)$ by a numerical integration rule:
$$
Z^N (s)  = \int_0^1 \Lambda (s_i)w_i, \; \; {\rm for} \; \;  w_i = s_{i-1}  - s_i, \; {\rm  or} \;  w_i = 1/2 ( s_{i-1} - s_{i+1} ) 
$$
In the case of nested sampling, one can use the deterministic grid points $s_i = \exp(-i/n) $ for  $i = 1,\ldots, m$ for a moderate $n$ and a large $m$. 
Polson and Scott (2014) recommends a modest $n = 20$ and a large $n = 1000$.

\paragraph{MC error} The advantage of this method is that the MC error bounds are $ O(N^{-4} ) $. For  $ p = \mathbb{I}_{ [0,1] } $, 
a convergence rate of $ 0( N^{-4} ) $ is available for a Riemann sum estimator of the form
$$
\hat{Z} = \sum_{i=1}^{N-1} ( u_{(i+1)} - u_{(i)} ) \frac{ L( u_{(i+1)} ) + L( u_{(i)} ) }{2}
$$ 
where $ u_{(i)} \sim U(0,1) $ are ordered uniform draws from $ p = \mathbb{I}_{ [0,1] }$. 
$$
\hat{Z} =  \sum_{i=1}^{N-1} ( x_{ (i+1) } - x_{ (i) } ) p( x_{(i)} ) L( x_{(i)} ) 
$$
where $ x_{(1)} \leq \ldots \leq x_{(n)} $ are the ordered sample associated with $(x_1, \ldots , x_n ) \sim P $.

\paragraph{Concentration Function}
 Let $P, Q$ be probability measures on $ x \in \mathcal{X} $. The likelihood ratio $ l_P(x) \defeq P(x) / Q(x) $ is a random variable whose expectation, 
 $E( l_P) = 1 $. 
 
 Let $ m(t) $ be the distribution function of $ l_P $ and 
 $$
 Q \left \{ x \in \mathcal{X} : l_P ( x ) \leq t \right \} = \int_{ l_P(x) \leq t } Q( x ) d x 
 $$
with  set-theoretic inverse
$m^{-1} ( x ) = \sup \left \{ t : m(t) \leq x \right \} $.
The concentration function is the Lorenz curve of the likelihood ratio r.v. defined by
$$
\phi_P ( u ) = \int_0^u m^{-1} ( z ) d z 
$$
Notice that $ m(0)=0 $ and $ m(1) =1 $

A weighted version of the concentration function can now be defined as follows.
First, define the titled measure
$$
Q^L( x ) \defeq \frac{ L(x) Q( x ) }{ \int_{\mathcal{X} }  L(x )Q( x ) d x }
$$
where now  the base measure $Q^L$ is chosen so that $  \int L(x) Q( x ) d x $ is known in closed form.

Now   $ m^L(0)=0 $ and $ m^L(1) = Z / Z_0 $  and we have  the following identity:
$$
 \frac{Z}{Z_0} = \frac{  \int_{\mathcal{X}} L(x) P(d x) }{    \int_{\mathcal{X}} L(x) Q(d x)  } 
$$
Hence a ratio of normalisation constants  can be written in terms of a concentration function.

\subsection{Nested Sampling}

Order states by their $x$-quantiles of the likelihood, $L$, under $\pi$.
Specifically, we assume $ z(\bx)$ has a well defined inverse  $ \bx(z) $ where 
$$
z(\bx) = Z_{L(\bx)}  =  \int_{L(\bx^\prime )> L(\bx) } \pi(\bx^\prime) d \bx^\prime
$$

Then under this change of variables, we can approximate the expectation of interest
via simple quadrature
$$
Z = \int_0^1 L( \bx(z) ) d z \approx \sum_{i=1}^N L_i ( x_i - x_{i-1} ) 
$$
One caveat is that the  $\bx(z)$ values near the origin contain all the mass.

The algorithm is as follows: if $ L_{\max} = \sup_{ \bx } \; L(\bx ) $ is known, we sample as follows:
set $ X=1, N=1, Z=0$
\begin{itemize}
\item Generate $ \bx^{(1)} \sim \pi(\bx) $ and set $L_0=0 , L_1 = L( \bx^{(1)} ) $.
\item If $ L_{\max} X < \epsilon Z $, then set  $Z=Z + (X/N) \sum_{j=1}^N L_{i+j-1} $ and stop.
\item Repeat while $ L_i X/N > \delta $
\item Generate $  \bx^{(i+N)} \sim \pi(\bx)\mathbb{I} \left ( L(\bx) > L_{i-1} \right ) $ and set $ L_i = L( \bx^{(i+N)} )$.
\item Set $N=N+1$ and sort $L_i$'s
\item Set $ Z=Z + L_i X/N $ and $N = N-1 , X = ( 1 - 1/N) X $.
\end{itemize}

\subsection{Split sampling} 

Split sampling uses importance sampling blankets, indexed by a weight function  $\omega(m)$ which can be adaptively estimated 
$$
g_\omega (\bx) = \frac{ \left \{ \int_0^{L(\bx)} \omega_s d s \right \} \pi(\bx) }{ \int_{\mathcal{X}} \Omega (\bx) \pi(\bx) d \bx } \; \;
{\rm where} \; \;  \Omega( L(\bx)) = \int_0^{L(\bx)} \omega_s d s \; .
$$
Suppose that $X \sim P(x)$ and $Y \equiv L(X)$.   Let $F_Y(y) = \mathbb{P}\{L(X) \leq y\}$ be the cumulative distribution function of $Y$.  Now define
\begin{equation}
\label{eqn:Zu}
Z(y) =1-F_Y(y) =  \int_{L(x) > y} d P(x)
\end{equation}
Clearly $Z(y)$ has domain $y \in \mathbb{R}^+$ and range $s \in [0,1]$ and is non-increasing in $y$.

Fubini interchange ``splits'' $Z$
$$
Z = \int_{\mathcal{X}} L(\bx) \pi(\bx) d \bx  = \int \int_{L(\bx)>m} \pi(\bx) d m = \int Z_m d m\; .
$$
\begin{itemize}
\item
Weight function $\omega(m)$. Marginal splitting density by
$$
\pi_{IS} \left ( m \right ) = \frac{ \omega(m)Z_m }{ \int_0^M \omega(m)Z_m d m } \; .
$$
Run MC,equilibrium $\pi(\bx)$ conditioned to the set $ \{ \bx : L(\bx) > m \} $ namely
$$
\pi ( \bx | m ) \sim \pi \left ( \bx | L(\bx) > m \right ) \; .
$$
\item
Auxiliary variable $m$: conditional doesn't depend on $Z_m$. 
$$
\pi ( m | \bx ) = \frac{ \omega(m) L_m(\bx) }{ \int_0^M \omega(m) L_m(\bx) d m } \; .
$$
\end{itemize}

\paragraph{Algorithm:} Draw samples $( \bx ,m)^{(i)} \sim \pi_{IS} \left ( \bx ,m \right )$ by iterating $ \pi_{IS} \left ( \bx |m \right ) $
and $ \pi_{IS} \left ( m|\bx \right )$. Estimate the marginal distribution $ \hat{\pi}_{IS} (m)$ via
$$
\hat{\pi}_{IS} (m) = \frac{1}{N} \sum_{i=1}^N \frac{ \omega(m) L_m(\bx^{(i)}) }{ \int_0^M \omega(m) L_m(\bx^{(i)}) d m } \; .
$$
The new estimate of the individual normalisation constants, $Z_m$, is
$$
\hat{Z}_m = \omega^{-1} (m)  \cdot \frac{\hat{\pi}_{IS} (m) }{ \hat{\pi}_{IS}(0) } \; .
$$
 Compute a new estimate, $\hat{Z}$, via
$$
\hat{Z} = \sum_{i=1}^N  \frac{ \Omega^{-1} \left ( L( \bx^{(i)} ) \right ) }{ \sum_{i=1}^N  \Omega^{-1} (L(\bx^{(i)})) } L(\bx^{(i)})    \; .
$$
We can use a discrete ``cooling'' schedule $ \omega(m) = \sum_{t=0}^T \omega_t \delta_{ m_t } ( m) $. 

Set $T=0$, $m_0 = 0$, $\Omega_0 = 1$, and $Z_0=1$.
While $T < T_{max}$, set $T=T+1$
\begin{itemize}
\item Simulate $\pi_{SS}(\mb{x})$ with $\{m_t\}_{0 \le t <T}$ and $\{\Omega_t\}_{0 \le t <T}$ until $N_{level}$ visits to level $T-1$.
\item Choose the $(1-\rho)$-quantile of likelihoods of level $T-1$ as $m_T$.
\item Set $\hat{Z}_T = \rho^{-T}$.
\item Set $\Omega_T = \hat{Z}_T^{-1}$.
\end{itemize}

Favor upper levels with condition (d1) Set $\Omega_T = e^{\Lambda T} Z_T^{-1}$. $\Lambda$ a boosting factor.  Time complexity to be $O(T)$.
Place more weight on the top level $T$ with condition  (d2) Set $\Omega_{T-1} = e^{\Lambda (T-1)} Z_{T-1}^{-1}$ and $\Omega_T = \beta \frac{e^{\Lambda T} -1}{e^\Lambda - 1} Z_T^{-1}$.

\paragraph{Discrete Split Sampling} is defined as follows: once we have found all levels, our split sampling algorithm runs:
\begin{itemize}
\item Set $i=0$ and $\nu_t = \nu_{init} \hat{Z}_t$ for each $t=0,1,\dots,T$. While $i \le n$, set $i=i+1$.
\begin{itemize}
\item Draws $\mb{x}^{(i)}$ using MH under weights $\Omega(m)$, and set $L_i = L(\mb{x}^{(i)})$.
\item Obtain $M^{(i)} = \Omega^{-1}(U_i)$ where $U_i \sim U(0,\Omega(L_i))$ with $\Omega(0)=1$.
\item For each $t$ with $m_t < L_i$, update $\nu_t = \nu_t + \Omega(L_i)^{-1}$.
\item Update $\hat{Z}_t = \nu_t/\nu_0$ and set $\Omega_t = \hat{Z}_t^{-1}$.
\end{itemize}
\end{itemize}
If $\omega(m)=1$ and  marginal, $ \mu_N(m) \equiv \hat{\pi}_{IS}(m) $ given by 
$$
 \mu_N (m) = \left \{ \frac{1}{N} \sum_{i=1}^N
\frac{\mathbb{I}{\{ m < L(\bx^{(i)}) \}}}{\Omega(L(\bx^{(i)}))} \right \} \omega(m) \; .
$$
Rebalance inversely proportional to visitation probabilities adjusts weights for $ m < \max_i L( \bx^{(i)} ) $
$$
\frac{ \omega^\star (m) }{ \omega(m) } = \frac{ \omega (m )}{ \mu_N(m) } = 
 \frac{ \mathbb{I}{\{ m < L(\bx^{(i)}) \}} }{ \Omega(L(\bx^{(i)}))} \; .
$$
 The ``flat histogram'' (FH) condition is as follows: for a tolerance $c$, 
$$
\max_{  m \in \{ m_t \} } \; \vline \; \mu_N ( m ) - \phi (m) \; \vline \; < c  \; .
$$
Update rule guarantees convergence in finite time, 
$$
\log \omega_{\kappa_N} ( m ) \leftarrow \log \omega_{\kappa_{N-1}} (m) + \gamma_{\kappa_N} \left ( \mu_{\kappa_N} ( m ) - \phi(m) \right ) \; .
$$

\section{Discussion}\label{sec:discussion}
The N-Queens completion problem is a conditional counting problem.  \cite{gent2017} shows that the N-queen completion problem if solved efficiently would imply that $P=NP$. In our framework, we need to compute 
$$
Z_O = E_{X \mid X = X_0}[S(x)] = \int_0^1 F^{-1}_{X \mid X = X_0}(u)du,
$$
where the expectation is taken with respect to the conditional distribution of $X$ given $X = X_0$. Here $X_0$ is the initial configuration of queens. The quantile re-ordering approach and the use of polynomial time sampling algorithms is still available -- we simply use the conditional density.










\bibliography{nqueens}
\end{document}